**Tools to help patients and other stakeholders' input into choice of estimand and intercurrent event strategy in randomised trials**

Joanna Hindley[1], Charlotte Hartley[1], Jennifer Hellier[2], Kate Sturgeon[1], Sophie Greenwood[3], Ian Newsome[4], Katherine Barrett[5], Debs Smith[5], Tra My Pham[1], Dongquan Bi[1], Beatriz Goulao[3], Suzie Cro[6], Brennan C Kahan[1]

[1] MRC Clinical Trials Unit at UCL, UCL, London, UK

[2] Department of Biostatistics and Health Informatics, Institute of Psychiatry, Psychology and Neuroscience, King's College London, London, UK.

[3] Aberdeen Centre for Evaluation, University of Aberdeen, Aberdeen, UK

[4] Public Partner with the MRC Clinical Trials Unit at UCL

[5] Public Parter with the National Institute for Health and Care (NIHR) Maudsley Biomedical Research Centre (BRC) at King's College London

[6] Imperial Clinical Trials Unit, School of Public Health, Imperial College London, London, UK

**Correspondence:** Brennan Kahan, b.kahan@ucl.ac.uk

**Abstract**

**Background**

Estimands can help to clarify the research questions being addressed in randomised trials. Because the choice of estimand can affect how relevant trial results are to patients and other stakeholders, such as clinicians or policymakers, it is important for them to be involved in these decisions. However, there are barriers to having these conversations. For instance, discussions around how intercurrent events (post-randomisation events which affect the interpretation or existence of the outcome) should be addressed in the estimand definition typically involve complex concepts as well as technical language. We aimed to provide tools that could facilitate conversations between researchers and patients and other stakeholders about the choice of estimand and intercurrent event strategy.

**Methods**

We developed three tools: (i) a video explaining the concept of an estimand and the five different ways that intercurrent events can be incorporated into the estimand definition; (ii) an infographic outlining these five strategies; and (iii) an editable PowerPoint slide which can be completed with trial-specific details to facilitate conversations around choice of estimand for a particular trial. Each tool was produced through collaboration between researchers and public partners.

**Results**

These resources can help to start conversations between the trial team and patients and other stakeholders about the best choice of estimand and intercurrent event strategies for a randomised trial. The video and infographic – which explain estimands and intercurrent events with reference to imagined examples – can be sent to stakeholders in advance of a consultation, or presented in the meeting itself. It is important that a member of the trial team is available to answer questions or clarify concepts following this. The editable slide can be completed by the trial team with the specific details of their trial, and then shown to patients or other stakeholders during the meeting to facilitate discussion around which intercurrent event strategy is most relevant for the trial. An example of a completed editable slide is also provided for an example weight loss trial.

**Conclusions**

We developed three tools to help researchers to have conversations with patients and other stakeholders about estimands, and how intercurrent events should be incorporated into the target estimand for a randomised trial.

**Key words:** estimand, intercurrent event, patient and public involvement, PPI, randomised trial

**Background**

Randomised trials are used to answer questions about how safe and effective treatments are, but for any given trial there are usually many different questions that could be addressed. For instance, a trial could evaluate the effect of a treatment if it were taken exactly as intended, or else, regardless of any treatment non-adherence. In order to clarify which question a trial is addressing, the use of estimands (which provide a precise description of the treatment effect to be estimated) are recommended.[1-8]

An important part of choosing an estimand is specifying which strategy will be used to handle intercurrent events. Intercurrent events are post-randomisation events that affect the interpretation or existence of outcome data. Common examples of intercurrent events include not receiving the assigned treatment, stopping the assigned treatment early, or receiving a different treatment to the one intended.[2,5,7] There are five different strategies that can be used to handle intercurrent events in the estimand definition: treatment policy, composite, while-on-treatment, hypothetical, and principal stratum. Details of these can be found elsewhere.[1,4]

The choice of intercurrent event strategy can impact the relevance of trial results for patients and other stakeholders, such as clinicians or policymakers.[4] For instance, an estimand which uses a hypothetical strategy to evaluate the effect of an intervention under perfect adherence may be less relevant for patients who are unlikely to be able to fully adhere. It is therefore important for patients and other stakeholders to be involved in these decisions for trials in which the results may inform their treatment choices.[9] Previous work has indicated that patients value being part of decisions around numerical aspects of trials[10] and in choices around estimands specifically.[9]

However, there are currently substantial barriers to involving patients and other stakeholders in discussions around the choice of estimand. The concepts around the different intercurrent event strategies can be challenging to understand, even to those with a statistical background, and much of the terminology and language is highly technical and may be inaccessible to many. Literature suggests that accessible language and communication tools can help to facilitate patient involvement in the more technical aspects of randomised trials, such as statistical aspects.[11] Cro *et al* have previously co-developed tools with public partners to introduce the concept of estimands[9] and explain the five attributes.[12] However, these do not introduce the strategies to handle intercurrent events, which are essential to the process of choosing an estimand.

We therefore co-developed tools with public partners that aim to facilitate discussions between members of the trial team and patients and other stakeholders about the choice of intercurrent event strategy to be used in the estimand definition. This paper outlines how these tools were developed, and how they can be used to facilitate discussions between researchers and stakeholders around the choice of estimand. This paper is primarily aimed at researchers, with the intent of describing how the tools were developed and how they could be employed within their own trials.

**Methods**

In conjunction with public partners, we co-developed three tools to help facilitate discussions between researchers and patients and other stakeholders around the choice of estimand and intercurrent event strategies, and to enable these stakeholders to better contribute to these decisions. The three tools are:

(i) A video explaining estimands and intercurrent events, and covering the five strategies that can be used to address intercurrent events in the estimand definition;
(ii) An infographic which describes each of these five intercurrent event strategies; and
(iii) An editable PowerPoint slide which can be completed with trial-specific details to facilitate conversations around choice of estimand for a particular trial.

*Video development*

The video focussed on describing estimands, intercurrent events, and the different intercurrent event strategies, using a fictional mental health trial as a motivating example.

We advertised through the Patient and Public Involvement team affiliated with the National Institute for Health and Care (NIHR) Maudsley Biomedical Research Centre (BRC) at King's College London for public partners to help us co-develop the video. Two public partners (KB, DS) were recruited. An initial meeting was held between several researchers and the public partners, to discuss the aims of the project and answer any questions.

An outline of the video was prepared by a team of researchers (CH, SG, TMP, JH, DB, KS, and BCK). Using this outline, a draft script was written by a science communications officer (CH). Feedback on the draft was sought from the research team and the public partners, and this was incorporated into a finalised script. The animation was developed by the science communications officer and feedback given by the research team. Adjustments were made based on this feedback and the video finalised.

*Infographic and editable slide*

The infographic focussed on describing intercurrent events and the different intercurrent event strategies. The editable slide focussed on enabling researchers to facilitate discussions with patients and other stakeholders around choice of strategy for a particular intercurrent event in a particular trial.

We advertised through the Patient and Public Involvement team affiliated with the MRC Clinical Trials Unit at UCL for public partners to help us co-develop these tools. One public partner (IN) was recruited. An initial meeting was held between one researcher (BCK) and the public partner, to discuss the aims of the project and answer any questions.

First drafts of the infographic, the editable PowerPoint slide, and a completed example of the editable slide were developed by members of a research team (BCK, SC, DD) and feedback solicited from the public partner. This was used to update and finalise the infographic, editable slide, and a completed example of the editable slide.

**Results**

*Video*

The video uses an imagined interaction between a doctor and a patient who is looking for a treatment for stress. The doctor looks at the results of a recently published randomised trial evaluating the use of cognitive behavioural therapy (CBT) for stress. A narrator explains how outcomes in this trial might depend on whether patients started taking anti-anxiety medication at some point during their course of CBT, and explains that this is an example of an intercurrent event. The narrator then explains that different research questions – or estimands – can be targeted depending on how this intercurrent event is addressed, and outlines the five different strategies for handling intercurrent events using the example trial. It is explained that the estimand should be chosen when the trial is being planned, and that not using an estimand may make the results of the trial difficult to interpret for patients and doctors.

The video is available online at: https://www.youtube.com/watch?v=UX8MmPo27rc

*Infographic*

The infographic is shown in Figure 1, and is available online at https://osf.io/sg4h7/, and with the supplementary material for this article.

Using an example of an imagined trial to evaluate a weight loss drug, it explains what estimands and intercurrent events are. Then, using an example of stopping treatment early as an intercurrent event, the five strategies for addressing this in the estimand definition are outlined for this trial, and it is emphasised that there is no 'right' estimand.

*Figure 1: An infographic explaining different ways that intercurrent events can be addressed in the estimands definition*

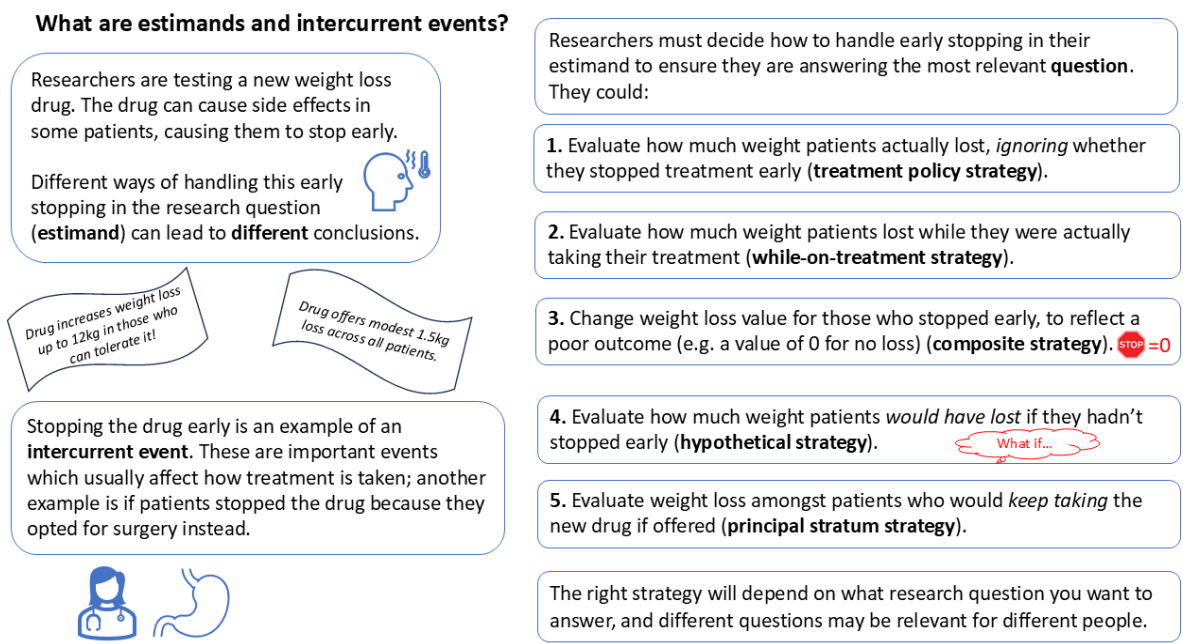

*Editable PowerPoint slide*

The editable slide is shown in Figure 2a, and is available online at https://osf.io/sg4h7/, and with the supplementary material for this article.

The highlighted fields show where researchers should update the slide with details of the randomised trial that they are planning. When completed, the slide will outline the different options that can be used to handle a specific intercurrent event in the planned trial.

An example of how this may be completed is shown in Figure 2b, and is available online at https://osf.io/sg4h7/, and with the supplementary material for this article.

This completed example uses a weight loss trial with the intercurrent event of stopping treatment early to show an example of how the editable slide could be completed.

*Figure 2a: An editable slide that can be used to illustrate the different approaches to addressing an intercurrent event in the estimand definition for a particular trial.*

**Which strategy should we use for [state intercurrent event] in the [TRIAL NAME] trial?**

We're undertaking the [trial name] trial to [give objective]. Some patients during the trial may [state intercurrent event, e.g. "stop treatment early", "switch to the alternate treatment", etc].

Different ways of handling this [state intercurrent event] in the research question (**estimand**) can lead to **different** conclusions.

[State intercurrent event] is an example of an **intercurrent event**. These are important events that occur during the trial which can affect how treatment is taken.

Insert trial logo here

We need to decide how to handle [state intercurrent event] in our estimand to ensure we are answering the most relevant **question**. We could:

**1.** Compare [outcomes] between treatments *ignoring* whether participants [experienced intercurrent event] (**treatment policy strategy**).

**2.** Compare [outcomes] before participants [experienced intercurrent event] (**while-on-treatment strategy**).

**3.** Change [outcome] values for those who [experienced intercurrent event], to reflect a poor outcome (e.g. a value of [give example value]) (**composite strategy**).

**4.** Compare [outcome] values patients *would have had* if they hadn't [experienced intercurrent event] (**hypothetical strategy**).

**5.** Compare [outcome] amongst patients who wouldn't [experience intercurrent event] under either treatment (**principal stratum strategy**).

There is no one right answer, and different questions may be relevant for different people. We want to know which question is most relevant for **you**.

*Figure 2b: An illustration of how the editable slide can be completed for an imagined weight loss trial.*

**Which strategy should we use for early treatment stopping in the WEIGHTLOSS trial?**

We're undertaking the WEIGHTLOSS trial to understand whether a new drug increases weight loss compared with placebo. Some patients during the trial may stop the new drug or placebo early.

Different ways of handling this early stopping in the research question (**estimand**) can lead to **different** conclusions.

Stopping the treatment early is an example of an **intercurrent event**. These are important events that occur during the trial which can affect how treatment is taken.

We need to decide how to handle early stopping in our estimand to ensure we are answering the most relevant **question**. We could:

**1.** Compare weight loss between treatments *ignoring* whether participants stopped early (**treatment policy strategy**).

**2.** Compare weight loss before participants stopped early (**while-on-treatment strategy**).

**3.** Compare weight loss after changing weight loss values for those who stopped early to a value of 0 to reflect a poor outcome (i.e. to denote no weight loss) (**composite strategy**).

**4.** Compare the weight loss values patients *would have had* if they hadn't stopped treatment early (**hypothetical strategy**).

**5.** Compare weight loss amongst patients who wouldn't stop early under either treatment (**principal stratum strategy**).

There is no one right choice, and different strategies may be relevant for different people. We want to know which question is most relevant for **you**.

### *Recommendations for using these resources*

Each of these tools takes a different approach to introduce intercurrent events and outline how these might be addressed in the estimand definition. The tools can be used individually or in combination with each other to facilitate conversations between a trial team and patients and other stakeholders. While they provide a starting point for these discussions, they do not give a comprehensive overview of the topic, and our aim is for them to form part of a wider consultation between members of the trial team and stakeholders around the choice of estimand for a trial. At these consultation meetings, at least one member of the trial team should have an understanding of estimands and the different intercurrent event strategies, so that they might answer any questions that arise.

The video and infographic provide a general introduction to estimands and intercurrent events. They can be presented during a meeting with stakeholders but can also be shared with them in advance to familiarise them with the topics to be discussed. Any questions from stakeholders can then be addressed at the meeting.

The editable slide shows more specifically how each approach to addressing intercurrent events in an estimand can look in the context of a planned clinical trial. It is intended to be used in a discussion aimed at gathering feedback on choice of estimand for a trial, involving members of a trial team with an understanding of estimands. It is therefore not intended for the slide to be shared with stakeholders in advance. When preparing the slide with the details of their planned trial, the trial team can consult our example shown in Figure 2b – this imagined example is not itself intended for presentation.

All of these tools examine approaches to including a single intercurrent event in an estimand definition. In reality, it is likely that multiple intercurrent events will occur in a trial, and each of these will require a different set of considerations when defining the estimand. This is a point that can be emphasised in discussions, and it could be useful to prepare multiple versions of the editable slide, one for each anticipated intercurrent event.

**Discussion**

It is vital that patients and other stakeholders, such as clinicians or policymakers, are given the opportunity to contribute to the choice of estimand to ensure that trials are addressing questions relevant to those using the results. However, this has so far been limited by complex nature of some concepts and the language used to describe these, together with the lack of tools to facilitate these conversations.

We developed three tools to address this gap, specifically focussing on how intercurrent events could be included in the estimand definition. These tools offer a general introduction to estimands and intercurrent events, and provide specific guidance on what different intercurrent event strategies look like in the context of the trial under consideration. Collectively and individually these tools can be used to initiate informed dialogue between the trial team and stakeholders.

These resources have the strength of being developed with input from public partners, who have given feedback on how they might be received in practice. They have not, however, been tested in real-world examples, and doing so might reveal further refinements or further need for patient and public involvement. While the video is fixed in its format, the infographic and editable slide can be updated directly by researchers for their planned trials. Another potential limitation of the tools is that we have prioritised brevity in each. While this suits the purpose of these as a precursor to - or part of - meetings with a trial team, it may also mean that some concepts are not immediately clear for all users. Finally, although clinical relevance is essential when choosing an estimand, it is not the only consideration. For instance, the ability to robustly estimate the chosen estimand is also important, and this should be considered alongside clinical relevance to inform the choice of estimand. As these tools do not address the fact that some intercurrent event strategies may be challenging to robustly estimate, further work around how these issues should be raised with stakeholders during discussions would be valuable.

**Conclusions**

We have developed three tools that can be used by research teams to facilitate discussions with patients and other stakeholders around how intercurrent events should be addressed in the estimand.

**Ethics approval and consent to participate**

Not applicable.

**Consent for publication**

Not applicable.

**Availability of data and materials**

All outputs from this work are available as part of the supplementary material of this manuscript, and online at the links provided within the manuscript.

**Competing interests**

None.

**Funding**

This work was supported by the MRC-NIHR Trials Methodology Research Partnership (MR/S014357/1). JH, KS, TMP, DB, and BCK are funded by the UK Medical Research Council (grant nos. MC_UU_00004/07 and MC_UU_00004/09).

**Authors' contributions**

BCK conceived of the study. JoH, CH, JeH, KS, SG, IN, KB, DS, TMP, DB, BG, SC, and BCK participated in the design of the study. JoH, CH, JeH, KS, SG, IN, KB, DS, TMP, DB, BG, SC, and BCK analysed and interpreted the data, revised the manuscript critically for important intellectual content, approved the final manuscript, and agreed to be accountable for its overall content.

**Acknowledgements**

We thank Darren Dahly for helpful conversations during the development of these tools.